\documentclass[prc,unsortedaddress,superscriptaddress,twocolumn,amssymb,showpacs,floatfix,nofootinbib]{revtex4-1}
 \usepackage{graphics}
 \usepackage{amsmath,amssymb}
 \usepackage{epsfig}
\usepackage{dcolumn}   

\begin{document}
\date{\today}
\title{An essential singularity of the cotangent of the Coulomb-nuclear phase shift, and a finite limit of the nuclear part 
  of the effective-range function derived at zero energy }
\author{Yu. V. Orlov}
\email{orlov@srd.sinp.msu.ru}
\affiliation{Skobeltsyn Institute of Nuclear Physics, Lomonosov Moscow State University, Leninskie gory, Moscow, Russia}

\date{\today}

\begin{abstract}
The Coulomb-nuclear phase shift $\delta^{(cs)}_l$, $\cot\delta^{(cs)}_l$ and a finite limit of the nuclear part
 $\Delta_l(k)$ of the effective-range  function (ERF) are derived for an arbitrary orbital momentum $l$ when energy $E\rightarrow0$.  It is proved that  $\cot\delta^{(cs)}_l$ has an essential singularity at zero energy, but $\Delta_l(k)$ does not. The explicit finite limit of $\Delta_l(0)$ is found. The property of $\Delta_l(k)$ as a meromorphic function makes  possible  the analytical continuation of a re-normalized scattering amplitude  from the physical energy region to a bound state pole. Then the asymptotic normalization coefficients (ANC)  can be deduced from experimental phase-shift data and applied to radiative capture processes which are important in  nuclear astrophysics for new elements creation. Our results are in agreement with  the  results published  for $S$ wave scattering.
\end{abstract}

\maketitle

\section{Introduction}
 Many reactions in supernovae explosions proceed through  bound states or low-energy resonance states.  One needs to find the asymptotic normalization coefficient (ANC) of the radial wave function  which can be used to calculate radiative capture cross sections. This process is one of the main sources of new element creation.

The conventional method of finding the ANC consists in fitting an experimental phase-shift $\delta^{(cs)}_l$ data presented by the effective-range function (ERF)  to  carry out an analytical continuation of the re-normalized (or effective) scattering amplitude from the physical region to a bound or resonance state pole. The  transformation of the  Coulomb-nuclear amplitude, which is not analytical,  to a re-normalized one is shown in the well-known paper Ref. \cite{Hamilton}. The re-normalized amplitude acquires  analytical properties similar to those of 
an the amplitude for a short-range interaction. 
  
Recently, a new $\Delta$ algorithm has been validated and applied to the ANC calculation in bound (Ref. \cite{OrlovIrgazievPRC96}) and resonance (Ref. \cite{IrgazievOrlovPRC98}) low-energy states instead of using the conventional effective-range function (EFR) method. In \cite{spren16}  the related form of the re-normalized scattering amplitude is proposed without proof. The $\Delta$ algorithm is much simpler than the ERF because it does not contain $h$, the Coulomb part of the ERF, including the psi-function. In fact, the function $h$   does not  appear in the re-normalized scattering amplitude. 

In contrast to the $\Delta$ method it was found  that the ERF method  is limited by the charge product of colliding particles. The $\alpha ^{12}$C system is  a good example of  such a situation when the nuclear part of the ERF is on average three orders of magnitude smaller than the Coulomb  part (see, for example, Ref. \cite{IrgazievOrlovPRC98}). This problem of the ERF method appears in  \cite{orlov16}
 where  the  ERF fittings for $\alpha ^{12}$C behave similarly in the different $^{16}$O states.  

However, in a number of publications (see \cite{Blokh}, \cite{LDlast}, \cite{sparenbergDelta}) the authors argue that the $\Delta$ method is not applicable to bound states since the Coulomb-nuclear re-normalized amplitude in this method does not allow an analytical continuation from the physical energy area to the region of negative energies. 
 The authors of  papers \cite{Blokh} and \cite{LDlast} claim that the $\Delta_l$ function (see Eq. (\ref{Delta})  below) supposedly has an essential singularity at zero energy. 

The main purpose of the present paper is to prove that the $\Delta_l(k)$ function has a finite limit at $E$=0. For the $S$ wave it was shown in the literature (see a special section below). 

   The re-normalized scattering amplitude has an explicit form because the only ingredient carrying  experimental information is the experimental Coulomb(c)-nuclear(s)  $\cot\delta^{(cs)}_l$ which is included in the  $\Delta_l$ function.  Analytic properties of  $\cot\delta^{(cs)}_l$ are very important for the ANC deducing. 
In the next section
 the $\Delta^{(cs)}_l$, $\delta^{(cs)}_l$ and $\cot\delta^{(cs)}_l$ are derived as the explicit functions of the relative momentum $k$ when $k a_B\rightarrow0$.  It is proved  that the $\Delta_l$ has a finite limit   when $E\rightarrow 0$. Thus the $\Delta_l(k)$ has no singularity at zero energy. This makes the $\Delta$ method no less mathematically strict than the  ERF method.

\section{Derivation of  the Coulomb-nuclear phase shift with its  cotangent and  the nuclear part of the ERF near zero energy }

 For 
 charge-less scattering,  the ERF is written as 
\begin{equation}\label{ERFstrong}
K^{(s)}_l(k^2)=k^{2l+1}\cot\delta^{(s)}_l=-\frac{1}{a^{(s)}_l}+\frac{1}{2} r^{(s)}_l k^2+...,
\end{equation}
where $a^{(s)}_l$ is the scattering length.
The standard effective-range expansion is used in (\ref{ERFstrong}), although   generally $K^{(s)}_l(k^2)$ is a meromorphic function.
So the Pad$\acute{e}$ approximant can be used instead of the effective-range expansion (ERE).
From Eq. (\ref{ERFstrong}) for $k\rightarrow0$ one gets
\begin{equation}\label{cots}
\cot\delta^{(s)}_l=-1/a^{(s)}_l k^{2l+1}.
\end{equation}
 
 Similarly to the procedure above, $\cot\delta^{(cs)}_l$, $\delta^{(cs)}_l$ near zero energy and at last $\Delta_l(0)$ are derived 
   from  the ERF for charged particles which
  has the following well-known form (see, for example, \cite{Haeringer}):
\begin{equation} \label{CoulombKl}
K_l(k^2) =2\xi D_l(k^2)\left[C_0^2(\eta)(\cot\delta^{(cs)}_l - i)+
 h(\eta )\right].
\end{equation}
 Here and below $l$ is the orbital momentum, $\xi=Z_1Z_2\mu\alpha$=1/$a_B$, $a_B$ is is the Bohr radius, $\eta=\xi/k$ is the Sommerfeld parameter, $k=\sqrt{2\mu E}$ is the relative momentum; $\mu$ and $E$  are the reduced mass and the center-of-mass  energy  of the colliding particles with  charge numbers $Z_1$ and $Z_2$ respectively; and $\alpha$ is the fine-structure constant. All formulas are given in the unit system $\hbar$=c=1. 

The following notations are used:
\begin{eqnarray}\label{CandDfunctions}
C_0^2(\eta)&=&\frac{\pi} {\exp(2\pi\eta)-1},\label{C02}\\
D_l(k^2)&=&\prod_{n=1}^l\Bigl(k^2+\frac{\xi^2}{n^2}\Bigr),\qquad D_0(k^2)=1 \label{DL_k}.
\end{eqnarray}
The $\Delta_l(k)$ is defined as
\begin{equation}  \label{Delta}
\Delta_l(k)=\frac{\pi\cot\delta^{(cs)}_l}{\rm{exp}(2\pi\eta)-\rm 1}.
\end {equation}

The Coulomb part is
\begin{equation} \label{H-eta-sigma}
h(\eta)= \psi(i\eta) + (2i\eta )^{- 1} - \ln(i\eta), 
\end{equation}
where $\psi(i\eta )$ is the digamma function.

 Using the explicit expression for
   $\psi(i\eta)$ in the form of the infinite sum, one can write
$h(\eta)$ as:
\begin{equation} \label{sum}
 h(\eta)=iC_0^2(\eta)+\eta^2\sum_{n=1}^\infty\frac{1}{n(n^2+\eta^2)}-\ln(\eta
)-\zeta,
\end{equation}
where $\zeta \approx$ 0.5772 is the Euler constant (see, for example, \cite{Arv74}).

Thus in the physical energy region, the ERF has a form (see (\ref{CoulombKl}) and (\ref{sum})) where its imaginary parts are mutually reduced:
\begin{equation} \label{CoulombKl-pos}
K_l(k^2) = 2 \xi D_l(k^2)\left[\Delta_l(k)+ \rm{Re}\it{h(\eta)}\right].
\end{equation}
 
In my opinion,  since $\Delta_l(k)$ and $\rm{Re}\it{h(\eta)}$ are two terms in the same expression in square brackets, the following statement is very controversial:  
 ``However, the validity of employing $\Delta_l(E)$ was not obvious since $\Delta_l(E)$, in contrast to $K_l(E)$, possesses an essential singularity at $E$ = 0'' (\cite{LDarxiv18} (page 2, section III, right column)). 

The limit of $\Delta_l(k)$ when $k\rightarrow0$ can be found directly from Eq. (\ref{CoulombKl-pos}) for the ERF in the physical region. 
In \cite{OrlovIrgazievPRC96} it is shown how the real  function Re$h(\eta)$  continues to the real function $h(\eta)$ in the region $E<0$.
 This function can be approximated by $1/12\eta^2$  around zero energy (see \cite{Hamilton} and Eqs. (6.3.18, 6.3.19) in 
\cite{abramowitz})  which has  equal limits for $E \to  0\pm$. 
Then   the Coulomb part in (\ref{CoulombKl-pos})
\begin{equation} \label{hfor0}
 \rm{Re}\it{h}\rm(\eta)\rightarrow0, \rm{when}\ \eta\rightarrow\infty \  or \ \it {E}\rightarrow \rm0.
\end{equation}
 
The function  $K^{(cs)}_l(k^2)$ can be expand as usual in series of powers $k^2$.
This leads to the relation
\begin{equation} \label{Coulomb not}
K^{(cs)}_l(k^2) = 2 \xi D_l(k^2)\Delta_l(k)=-1/a^{(cs)}_l,
\end{equation}
where only the first term is taken into account when $k\rightarrow0$ in the ERE on the right side of the equation.
After taking the limit of the $D_l{(k^2)}$ when $ka_B\rightarrow0$ (see Eq. (\ref{DL_k}),
\begin{equation} \label{D in Zero}
D_l(k^2)\rightarrow\xi^{2l} /(l!)^2,
\end{equation}
 the following limiting relations are obtained:
\begin{eqnarray}\label{finiteLimit}
\Delta_l(k) \rightarrow \Delta_l(0) = \frac{-(l!)^2} {2a^{(cs)}_l\xi^{2l+1}}, \label{l>0}\\
\Delta_0(0)=-\frac{a_B}{2a^{(cs)}_0} \label{l=0}
\end{eqnarray}

As a result, the absence of the essential singularity in the $\Delta_l(k)$ function at zero energy is  proved. The $\Delta_l(k)$ has a finite explicit  limit which depends on $l$, $\xi$ and the scattering length $a^{(cs)}_l$.  Eqs. $(\ref{l>0})$ and  $(\ref{l=0})$  define the new  relationship  between $\Delta_l(0)$ and the scattering length $a^{(cs)}_l$. For the Pad$\acute 
e$ approximation, a combination of fitted constants appears instead of $a^{(cs)}_l$.

Next the explicit expressions for  $\cot\delta^{(cs)}_l$ and $\delta^{(cs)}_l$ are derived.
The following formula  is obtained from the definition of $\Delta_l(k)$ Eq. (\ref{Delta}) and from the last relation
 Eq. (\ref{finiteLimit}) when $ka_B\rightarrow0$ and $l>0$:
\begin{equation}\label{cotangens}
\cot\delta^{(cs)}_l(k)=-\frac{(l!)^2}{2\pi\xi^{2l+1} a^{(cs)}_l}\left[\exp\left(\frac{2\pi}{k a_B}\right)-1\right].
\end{equation}
One  can ignore 1 in square brackets and get the final formulas:
\begin{eqnarray}\label{cotdelta}
\cot\delta^{(cs)}_l(k)=-\frac{(l!)^2}{2\pi\xi^{2l+1} a^{(cs)}_l}\;\rm{\exp}\left(\frac{2\pi}{\it{k\:\!a_B}}\right),\label{cotdelta0}\\
\delta^{(cs)}_l(k)=-\frac{2\pi}{(l!)^2}\xi^{2l+1}a^{(cs)}_l \rm{\exp}\left(-\frac{2\pi}{\it{k\:\!a_B}}\right),\label{delta0}
\end{eqnarray}
when  $\eta\gg1$ or $ka_B\it \ll\rm{1}$ ($\cot\delta\rightarrow1/\delta$ when $\delta\rightarrow0$).
Using (14), the following expressions are derived:

\begin{eqnarray}\label{cotdelta}
\cot\delta^{(cs)}_0(k)=-\frac{1}{2\pi\xi a^{(cs)}_0}\;\rm{\exp}\left(\frac{2\pi}{\it{k\:\!a_B}}\right),\label{scotdelta}\\
\delta^{(cs)}_0(k)=-{2\pi}\xi a^{(cs)}_0 \rm{\exp}\left(-\frac{2\pi}{\it{k\:\!a_B}}\right) \label{sdelta}.
\end{eqnarray}

So, the factor in the denominator of Eq. (\ref{Delta}), which contains the essential singularity, is reduced with the same factor of $\cot\delta_l$ Eq. (\ref{cotdelta0})   in the nominator of Eq. (\ref{Delta}) when $E\rightarrow0$. 
 This leads to the formulas (\ref{finiteLimit}) and (\ref{l=0}) for the $\Delta_l(k)$ function when $ka_B\rightarrow0$.
Eqs. (\ref{finiteLimit})  to  (\ref{sdelta})   complete our derivation of the formulas for $\Delta_l(0)$, $\cot\delta^{(cs)}_l(k)$ and  $\delta^{(cs)}_l(k)$ near  zero energy.

\section{Agreement of the obtained results with  information from the literature}

The related expression for the limit of $\delta^{(cs)}_l(k)$:
\begin{equation} \label{delta zero energy}
\delta^{(cs)}_l(k)=-\frac{2\pi}{(l!)^2}\xi^{2l+1}a^{(cs)}_l \rm{\exp}\left(-\frac{2\pi}{\it{ka_B}}\right),
\end{equation}
when  $\eta\gg1$ or $a_B\it k\ll\rm{1}$, 
 is presented in Ref. \cite{mur83} (Eq. (1.5)) but its derivation is not given.

The Coulomb-nuclear phase shift $\delta^{(cs)}_l$ is defined for a strong ($s$) short-range interaction  in the presence of the Coulomb ($c$) force, while $\delta^{(s)}_l$ is defined for a strong ($s$)  interaction. It is known that they are quite different functions of  momentum 
$k$. Eq. (\ref{delta zero energy}), which coincides with the derived above Eq. (\ref{delta0}), shows that the $\delta^{(cs)}_l(k) \rightarrow0$ when $k\rightarrow0$, while the $\delta^{(s)}_l(k)\rightarrow n\pi$ (n is an integer). The latter limit is related to the Levinson's theorem. 
Due to this difference between the phase shifts $\delta^{(cs)}_l$ and $\delta^{(s)}_l$, which is stressed in Ref. \cite{LandSmorod}, the Levinson's theorem is applicable only to the phase shifts $\delta^{(s)}_l$ for a short-range interaction and  not to the phase shift $\delta^{(cs)}_l$.

As it is noted in Refs. \cite{OrlovIrgazievPRC96} and \cite{IrgazievOrlovPRC98} and deduced in the previous section, the presence
 of the essential singularity in $\cot\delta^{(cs)}_l$ compensates for the same singularity in the denominator of the $\Delta_l$ function.  
 Therefore   an analytical continuation of the re-normalized scattering amplitude from the physical region to the negative energy is  mathematically  rigorous.

 Our general results for the arbitrary $l$ agree with those in the section  ``Resonance scattering  of  charged  particles'' of Ref. \cite{LandSmorod} where   the $S$-wave resonance near a threshold is studied. There the related expression  is presented  for $\cot\delta^{(cs)}_0$ 
(138.11), taking into account only the first term in the ERF expansion. We note here only the presence of the factor $\exp(2\pi\eta)$ which confirms the essential singularity of $\cot  \delta^{(cs)}_0$.
It is peculiar that a typo in Eq. (138.12)) in the American third edition (1977)  repeats that in Eq. (136.12) in the second Russian edition (1963):
 the exponent in the denominator is written as $2\pi k a_c$ instead of $2\pi/k a_c$ (in our notation $a_c$ = $a_B$).

Last but not  least, support to our results is given in  subsection (4.3) ``Effective-range theory for proton-proton scattering'' in book \cite{Brown} for the $S$-wave. The main conclusions, supported by the formulas that are written below in our notations, are as follows.

 The function
\begin{equation}\label{BrownERF} 
f(E)=\Delta_0(\eta)+\rm {Re} \it{h(\eta)}
\end{equation}
should have a limit when $E\rightarrow0$ or $\eta\rightarrow\infty$. Indeed, as is shown in Ref. \cite{tromborg}, $f(E)$ is a regular function of $E$ for the  Yukawa potential superposition in the  half-plane Re$E>$0 and has a cut along the negative semi-axis when Re$E \leq \:E_d$. $E_d$ is the negative energy which connected to the radius of nuclear forces. Due to this, $f(E)$ can be expanded into a Taylor series around the point $E$=0. The limit of $f(E)$ when $E\rightarrow0$ is defined as
\begin{equation}\label{BrownDelta0}
f(E=0)=[\Delta_0(\eta)+\rm {Re} \it{h(\eta)}]\mid_{E=\rm{0}}=-\frac{a_B}{\rm{2} \it{a_p}},
\end{equation}
where $a_p$=$a^{(cs)}_0$ is the proton-proton scattering length.
Thus, we obtain
\begin{equation}\label{BrownERE}
\frac{\Delta_0(\eta)+\rm {Re} \it{h(\eta)}}{a_B}\approx-\frac{1}{a_p}+\frac{k^2 r_0}{2}-Pk^4r^3_0+...
\end{equation}
 Formula (\ref{BrownERE}) completes the  citation from book \cite{Brown}.
The results outlined above have predecessors that have begun to study the problems under consideration (see, for example, Ref. \cite{LandauBethe}).
Since Re$\;h$=\;0 at $E$= 0, Eq. (\ref{BrownDelta0}) coincides with Eq. (\ref{l=0}) for $l$= 0. This is a direct proof for the essential singularity absence in the $\Delta_0(\eta)$. 

 Note, that in (4.22) $h(\eta)$= Re\;$h(\eta)$ in our notations.  Eq. (\ref{hfor0}) coincides with (4.1 7b) in \cite{Brown}). There are no objections to this formula in the literature.
 
The typos in this subsection of Ref. \cite{Brown} should be corrected.  In $C^2_0$ definition (4.16) $[2\pi\eta/(\rm{exp}(2\pi\eta)-1)]^2$ should be replaced by $2\pi\eta/(\rm{exp}(2\pi\eta)-1)$ and in (4.23) $C^2$ should be replaced by
$C^2_0$.

It follows from the above that the $\Delta_0(\eta)$ is also a meromorphic function and therefore can be expanded into a Taylor series or represented by a Pad$\acute{\tilde}{e}$-approximant. There is no  obstacle for the exclusion of the function Re$h(\eta)$ from  Eq. (\ref{BrownERE}).   After that, the coefficients in the right part of  Eq. (\ref{BrownERE}) will change.   Re\;$h(\eta)$ is not included in the expression for the re-normalized scattering amplitude, as explained in detail in  Ref. \cite{IrgazievOrlovPRC98}.

Thus, the  equations in the sections II and III, especially Eq. (\ref{BrownDelta0}), validate our conclusion that there  is no   $\Delta_l(k)$ essential singularity at zero energy. Otherwise not only $\Delta_l(k)$ but the whole ERF function could not be considered as a meromorphic function at low energies and used to find the ANC. 

\section{conclusion}

In the present paper the explicit expressions of the Coulomb-nuclear phase shifts, their cotangents and the $\Delta_l(k)$ function are derived using the ERF well-known formula.  
A finite explicit formula for $\Delta_l(0)$ is obtained which  clearly depends on $l $, $a_B$ and the scattering length. This means  that $\Delta_l(0)$ has no essential singularity and there are  no obstacles to applying the $\Delta$ method to deducing ANCs from an experimental phase-shift input.

The results of the present paper are in agreement with the literature including books  considering the $S$-wave state. In book \cite{Brown} for the proton-proton system an equation is given for $\Delta_0(0)$ which coincides with  Eq. (15) derived in our work. The related results validates  that the ERF is a meromorphic function and support the main point of the $\Delta$ method that the $\Delta_0(\eta)$ function is also meromorphic. Therefore $\Delta_0(\eta)$ can be expanded into a Taylor series or represented by a Pad$\acute{e}$-approximant.

One needs to highlight a reference to Levinson's theorem in  Ref. \cite{LDlast} for systems with two bound states with the same $J^{\pi}$ on the example of the nucleus $^{16}$O. The phrase ``The pole at $E > 0$ is due to the Levinson theorem.'' from Ref. \cite{LDlast} (see page 024602-5, left hand column in front of the subsection A) may indicate a possible source for the incorrect statement about the essential singularity of the $\Delta_l$ function. This theorem is valid only for a pure strong interaction without Coulomb repulsion. If $\cot\delta^{(cs)}_l$ in the nominator of (\ref{delta zero energy})  is changed to $\cot\delta^{(s)}_l$, then the essential singularity in $\Delta_l$
indeed will occur because of the denominator.

In \cite{OrlovIrgazievPRC96} and \cite{IrgazievOrlovPRC98}, it is stated that the singularity in  $C_0^2$ is compensated by the corresponding singularity in  $\cot\delta^{(cs)}_l(k)$,  due to the Coulomb-nuclear phase-shift $\delta^{(cs)}_l(k)$  behavior near zero (see (\ref{delta zero energy})). 

 Thus, the claim in \cite{Blokh} and \cite{LDlast}  about the supposed 'essential $\Delta_l$ singularity' is incorrect, as is the criticism of the Delta method which is formulated, validated and used in Ref. \cite{OrlovIrgazievPRC96}.  Note that   book \cite{book} includes a correct short description of the $\Delta$ method.

 In sum, the present paper proves that  there is no essential singularity in the $\Delta_l(k)$ function at zero energy. The $\Delta_l(0)$ finite explicit expression  is found.  In addition, Eq. (\ref{finiteLimit}) can be used to estimate the scattering length when the ERF method becomes invalid. If the scattering length is known then $\Delta_l(0)$ values can be included in the input experimental data for better fitting.
 The main conclusion of the present paper:  there are no obstacles to  applying the $\Delta$ method to deducing the ANC values, which are important for astrophysics and for the direct reaction theory using Feynman diagrams.

\section*{ACKNOWLEDGEMENTS}
This work was partially supported by the Russian Science Foundation (Grant No.16-12-10048). 
The author is grateful to A.E. Kudryavtsev  for a useful discussion and to H.M. Jones for editing the English of this manuscript.


\begin{thebibliography}{00}
		\bibitem{Hamilton} J.~Hamilton, I. ~{\O}verb\"o, and B.~Tromborg, Nucl. Phys. B \textbf{60}, 443 (1973).
		\bibitem{OrlovIrgazievPRC96} Yu.\,V.~Orlov, B.\,F.~Irgaziev, and Jameel-Un Nabi, Phys. Rev. C \textbf{	96}, 025809 (2017); arXiv:1702.04933.
	\bibitem{IrgazievOrlovPRC98}  B.\,F.~Irgaziev, Yu.\,V.~Orlov,  Phys. Rev. C \textbf{98}, 025809 (2017)	; arXiv:1801.05933.

		\bibitem{spren16} O.\,L.~ Ram\'irez Su\'arez, and J.\,-M.~Sparenberg, Phys. Rev. C \textbf{96}, 034601 	(2017); arXiv:1602.04082.
		\bibitem{orlov16} Yu.\,V.~Orlov, B.\,F.~Irgaziev, and L.\,I.~Nikitina, Phys. Rev. C \textbf{93}, 014612 	(2016);  Phys. Rev. C \textbf{93}		, 059901(E) (2016).
	\bibitem{Blokh} L.\,D.~Blokhintsev, A.\,S.~Kadyrov, A.\,~M. Mukhamedzhanov, and D.\,A.~Savin, arXiv:1710.10767.
	\bibitem{LDlast} L.\,D.~Blokhintsev, A.\,S.~Kadyrov, A.\,~M. Mukhamedzhanov, and D.\,A.~Savin, Phys. Rev. C \textbf{97}, 024602 (2018).
		\bibitem{sparenbergDelta} D.\,~Gaspard and J.\,-M.~Sparenberg, Phys. Rev. C \textbf{97}, 044003 (2018).
		
	

		\bibitem{Haeringer} H.\, Van Haeringen, J. Math. Phys., \textbf{18}, 927 (1977).
		\bibitem{Arv74} J. Arvieux, Nucl. Phys. A\textbf{221}, 253 (1974).
		\bibitem{LDarxiv18}  L.\,D.~Blokhintsev, A.\,S.~Kadyrov, A.\,~M. Mukhamedzhanov, and D.\,A.~Savin , arXiv:1810.06812. 
		\bibitem{abramowitz} \it{Handbook of Mathematical Functions with Formulas, Graphs and Mathematical Tables},\rm{ Edited by M. Abromowitz and I. A. Stegun [National Bureau of Standards, Applied Mathematics Series-55, (1964)]}.
		
		\bibitem{mur83} V.\,D.~Mur, A.\,E.~Kudryavtsev, and V.\,S.~Popov, Yad. Fiz., \textbf{37}, 1417 (1983) [Sov. J. Nucl. Phys., \textbf{37},
		844 (1983)].
		\bibitem{LandSmorod} L.\,D.~Landau and E.\,M.~Lifshits, \it{Quantum Mechanics, Non-relativistic theory.} \rm{Third edition. Pergamon Press.} 	
	
	\bibitem{Brown} G.\,E.~Brown, and A.\,D.~Jackson, \it{The nucleon-nucleon interactions}, \rm{North-Holland Publishing Company. 1976.}
	 \bibitem{tromborg} B. Tromborg, and J. Hamilton, Nucl. Phys. B \textbf{76}, 483 (1974).
	\bibitem{LandauBethe} L.\,D. ~Landau and J.\,A. ~Smorodinsky, J. Phys. Acad. USSR \textbf{8}, 154 (1944) ;
H.\,A. ~Bethe, Phys. Rev. 76 (1949) 38.
	
		
	\bibitem{book} Leonid Blokhintsev, Yuri Orlov, and Dmitri Savin, \it{ Analytic and Diagram Methods in Nuclear Reaction Theory}
	\rm(Nova Science Publishers, Inc., New York, 2017).
 
		
\end{thebibliography}
\end{document}